\documentclass{jfm}
\usepackage{graphicx}
\usepackage{epstopdf, epsfig}
\usepackage{booktabs}
\usepackage{float}
\newcommand{\be}{\begin{equation}}
\newcommand{\ee}{\end{equation}}
\newcommand{\eps}{\epsilon}

\graphicspath{{./figures/}}

\usepackage{latexsym,amsmath,amssymb,graphicx,booktabs}
\usepackage{cleveref}
\usepackage{color}
\usepackage{multicol}
\usepackage{chngcntr}

\makeatletter
\newcommand{\specialnumber}[1]{%
	\def\tagform@##1{\maketag@@@{(\ignorespaces##1\unskip\@@italiccorr#1)}}}
\newcommand{\specialeqref}[2]{\begingroup
	\def\tagform@##1{\maketag@@@{(\ignorespaces##1\unskip\@@italiccorr#2)}}%
	\eqref{#1}\endgroup}

\shorttitle{Stochastic transport by breaking waves}
\shortauthor{D. Eeltink, R. Calvert, J.E. Swagemakers, Qian Xiao and T.S. van den Bremer}

\title{Stochastic particle transport by deep-water irregular breaking waves}

\author{D. Eeltink\aff{1,2,3},
  R. Calvert\aff{4,5},
  J.E.  Swagemakers\aff{4}, Qian Xiao\aff{2}
 \and T.S. van den Bremer\aff{2,4}\corresp{\email{t.s.vandenbremer@tudelft.nl}}}
\affiliation{\aff{1}Department of Mechanical Engineering, Massachusetts Institute of Technology, Cambridge, Massachusetts, USA
\aff{2}Department of Engineering Science, University of Oxford, Parks Road, Oxford OX1 3PJ, UK
\aff{3} Laboratory of Theoretical Physics of Nanosystems, EPFL, Ch-1015 Lausanne, Switzerland 
\aff{4}Faculty of Civil Engineering and Geosciences, Delft University of Technology, 2628 CD, Delft, The Netherlands
\aff{5}School of Engineering, University of Edinburgh, Edinburgh, EH9 3FB, UK}

\begin{document}
\maketitle
\begin{abstract}

Correct prediction of particle transport by surface waves is crucial in many practical applications such as search and rescue or salvage operations and pollution tracking and clean-up efforts. Recent results by \citet{Deike2017} and \citet{Pizzo2019} have indicated transport by deep-water breaking waves is enhanced compared to non-breaking waves. To model particle transport in irregular waves, some of which break, we develop a stochastic differential equation describing both mean particle transport and its uncertainty. The equation combines a Brownian motion, which captures non-breaking drift-diffusion effects, and a compound Poisson process, which captures jumps in particle positions due to breaking. From the corresponding Fokker--Planck equation for the evolution of the probability density function for particle position, we obtain closed-form expressions for its first three moments. We corroborate these predictions with new experiments, in which we track large numbers of particles in irregular breaking waves. For breaking and non-breaking wave fields, our experiments confirm that the variance of the particle position grows linearly with time, in accordance with Taylor’s single-particle dispersion theory. For wave fields that include breaking, the compound Poisson process increases the linear growth rate of the mean and variance and introduces a finite skewness of the particle position distribution.
\end{abstract}


\section{Introduction}
Correctly understanding and predicting the motion of objects on the ocean surface is crucial for several applications. For example, floating plastic marine litter has rapidly become an acute environmental problem (e.g.,  \cite{Eriksen2014}). A mismatch exists between the estimated amount of land-generated plastic entering coastal waters ($5$-$12$ million tonnes yr$^{-1}$, \cite{Jambeck2015}) and the estimated total amount of plastic floating at sea (less than $0.3$ million tonnes, \cite{Cozar2014,Eriksen2014,vanSebille2015}), which necessitates more accurate transport and dispersion models \citep{vanSebille2020}. In addition, efforts to clean up floating plastics rely on an accurate prediction of the distribution and trajectories of particles to deploy clean-up devices at the right location (e.g., \cite{SainteRose2020}). Similarly, estimating environmental impact for oil spills and open-sea rescue operations rely crucially on transport models (e.g., \citet{Christensenetal2018}).

During the periodic motion of a (deep-water) surface gravity wave, a fluid parcel does not follow a perfectly circular trajectory. Instead, it experiences a net drift in the direction of wave propagation, known as the Stokes drift \citep{stokes1847}. Wave models such as WAM and WaveWatch III predict wave fields properties averaged over longer timescales (typically 3 hours), based on which estimates of the mean Stokes drift over that period can be made \citep{Webb2011, Breivik2014}. This Stokes drift prediction is typically superimposed onto a Eulerian flow field, often given by ocean general circulation models or measurements. A number of recent studies have shown that the inclusion of the Stokes drift in this way can alter the predicted direction of plastic pollution transport, shifting convergence regions  \citep{Dobler2019} and pushing microplastics closer to the coast \citep{Delandmeter2019, Onink2019}.

However, for several reasons, the Stokes drift alone should not be the velocity with which waves actually transport objects on the ocean surface. First and foremost, it is the wave-induced Lagrangian-mean velocity, made up from the sum of the Stokes drift and the wave-induced Eulerian-mean velocity, with which waves transport particles. On the rotating Earth, the Coriolis force in combination with the Stokes drift drive an Eulerian-mean current in the turbulent upper-ocean boundary layer, known as the Ekman--Stokes flow, which includes the effect of boundary-layer streaming \citep{longuethiggins1953}. 
This Ekman--Stokes flow needs to be added to the Stokes drift in order to predict the wave-induced Lagrangian-mean flow with which particles are transported \citep{Higgins2020}, or wave and ocean circulation models need to be properly coupled (e.g., \citet{Staneva_etal2021}). The Ekman--Stokes flow can have significant consequences for global floating marine litter accumulation patterns \citep{Cunningham2022}. Wave-induced Eulerian-mean currents can also play a role in laboratory experiments \citep{VanDenBremer2019,Bremer2018}. Second, the properties of the object itself such as its shape, size and buoyancy can cause the object's trajectory to be different from that of an infinitesimally small Lagrangian particle with a different mean transport as a result \citep{santamaria_etal2013,Huang2016, Alsina2020,Calvert2021,DiBenedetto_etal2022}. Third, breaking waves are known to transport particles much faster than non-breaking waves. For steep waves, particles may surf on the wave \citep{Pizzo2017} and be subject to transport faster than the Stokes drift due to wave breaking \citep{Deike2017,Pizzo2019}. In this paper, we will focus on (almost) perfect Lagrangian particles and consider the influence of unidirectional irregular deep-water waves and wave breaking, thus ignoring the effects of Coriolis forces, wind and non-wave-driven currents, which evidently also determine the drift of an object in the real ocean.

For non-breaking waves, the Stokes drift can be estimated estimated for various sea states (e.g., \citet{Webb2011, Breivik2014}) with an important but not always resolved role for the spectral tail \citep{LenainPizzo2020}. In an irregular or random wave field, the Stokes drift becomes a stochastic process itself. That is, due to the random wave field, particles will diffuse with respect to this mean velocity, yielding a distribution in their predicted position \citep{Herterich1982}. Estimates of the variance of the Stokes drift can either be obtained from the spectrum \citep{Herterich1982} or from the joint distribution of the significant wave height and peak period \citep{Longuet-Higgins1983,Myrhaug2014}. Alternatively, deterministic simulations of the particle trajectories can be performed (e.g.,  \cite{Li2021,Farazmand2019}), which could then allow for calculation of statistics of the evolution for different initial conditions using Monte Carlo methods.

The particle transport of breaking waves can be described at different scales. At the scale of individual waves or wave groups, deterministic models for particle trajectories can be formulated to include wave breaking. Direct numerical simulations of breaking focused wave groups show a linear scaling of the net particle transport with the theoretical steepness of the wave group at the focus point $\mathcal{S}$ \citep{Deike2017}, in contrast to the quadratic scaling of Stokes drift with steepness for non-breaking waves. \citet{Restrepo2019} confirm the linear scaling with $\mathcal{S}$ and calculate the variance of the drift. Experimental confirmation of the enhanced drift and the linear scaling with steepness for wave groups is provided in \cite{Lenain2019} and \cite{Sinnis2021}, where \citet{Sinnis2021} also consider the effects of bandwidth. Considering much larger scales, relevant for application to the real ocean, requires a stochastic approach. \cite{Pizzo2019} extend the result obtained by \cite{Deike2017} for wave groups to sea states. Based on the wave spectrum (peak wavenumber) and wind speed, one can estimate the breaking statistic $\Lambda(c){\rm d}c$ \citep{Banner2000,Dawson1993,Ochi1983,Holthuijsen1986,Sullivan2007}, defined as the average length of breaking crests moving with a velocity in the range $(c,c+{\rm d}c)$, where $c$ is the phase speed \citep{Phillips1985}. Consequently, the breaking drift speed found in \cite{Pizzo2019} is weighted by the percentage of broken sea surface per unit area, which in turn can be computed as a function of peak wavenumber and wind speed. Comparing their prediction of the drift speed to the Stokes drift for non-breaking waves shows that, as the wind velocity and wave steepness increase, the wave breaking component of drift becomes more important and can be as large as 30\% of the Stokes drift for non-breaking waves \citep{Pizzo2019}.

A series of papers have examined stochastic Stokes drift \citep{Jansons1998,Bena2000}, focusing on diffusion in the case of two opposing waves, for which the mean drift should be zero, yet diffusion still causes transport.
Their insights cannot immediately be applied to realistic ocean waves. More generally, several authors have examined the Taylor particle diffusion of a random surface gravity wave field \citep{Herterich1982,Sanderson1988,Weichman2000,Balk2002,Balk2006}. \citet{Buhler2009} derive the Taylor single-particle diffusivity for random waves in a shallow-water system under the influence of the Coriolis force and corroborate their results with Monte Carlo simulations. A generally applicable stochastic framework is provided in \cite{Buhler2015}, who derive a stochastic differential equation (SDE) for the particle position and a Kolmogorov backward equation for particles along quasi-horizontal stratification surfaces induced by small-amplitude internal gravity waves that are forced by white noise and dissipated by nonlinear damping designed to model attenuation of internal waves.  

In this paper, we propose a stochastic differential equation (SDE) for the evolution of particle position in a unidirectional irregular deep-water wave field with wave breaking and obtain the corresponding Fokker--Planck equation for the evolution of the distribution. The SDE combines a Brownian motion, which captures non-breaking drift-diffusion effects, and a compound Poisson process, which captures jumps in particle positions due to breaking. We focus on the short-time regime over which the properties of the sea state (i.e., its spectrum) stay constant and corroborate the predictions of our SDE with new laboratory experiments in which we track a large number of particles.

The paper is organized as follows. First, \S \ref{sec:model} introduces the Brownian drift-diffusion process to model particle position evolution without breaking and the Poisson process to model the surfing behavior experienced when a particle encounters a breaking crest. The corresponding Fokker--Planck equation for the evolution of the probability density function of particle position and its mean, variance and kurtosis are also derived in \S \ref{sec:model}. Then, \S \ref{sec:experiment} outlines the wave basin experiments performed, where particles were tracked in irregular waves. In \S \ref{sec:results}, we corroborate our theoretical predictions with the experimental results for different wave steepnesses and, consequently, different fractions of breaking waves. Finally, we conclude in \S \ref{sec:conclusion}.

\section{Stochastic model for particle transport}\label{sec:model}
For a given initial particle position $X_0=X(t=0)$, we seek to determine the (long-term) evolution of the particle position  $X(t)$ and its distribution, where we will only consider wave-averaged (Eulerian-mean or Lagrangian-mean) quantities. Particles are transported with the Lagrangian-mean velocity, which  consists of the sum of the Eulerian-mean velocity and the Stokes drift (e.g., \cite{buhler2014}). In our case, we consider the Lagrangian-mean velocity:
\begin{equation}
    u_\text{L} = u_\text{E,NB} + \underbrace{u_\text{S} + u_\text{B}}_{u_\text{D}},
    \label{eq:Lagrangian-mean}
\end{equation}
which consists of the sum of the Eulerian-mean velocity that excludes the effects of wave breaking $u_\text{E,NB}$ and a drift velocity $u_\text{D}$, which in turn consists of the Stokes drift for non-breaking waves $u_\text{S}$  and a breaking contribution $u_\text{B}$.  For simplicity, we subsequently ignore the wave-induced Eulerian-mean velocity $u_\text{E,NB}$ in our model, as its contribution to drift on the surface of the (non-rotating) ocean is negligible for deep-water waves (e.g., \cite{vandenbremer_taylor2015,Higgins_etal2020}). To compare to basin experiments (in \S \ref{sec:experiment}), we will take the effect of Eulerian flow (wave-induced or otherwise) into account. 

We propose to model the (wave-averaged) particle position $X(t)$ as a jump-diffusion process for which the stochastic differential equation (SDE) can be written as:
\begin{align} \label{eq:JumpDiffSDE}
           {\rm d}X &=& \underbrace{b(X(t),t) {\rm d}t}_{\text{Mean drift}}  \hspace{0.5cm}&+& \underbrace{\sigma (X(t),t) {\rm d}W(t) }_{\text{Diffusion process}}\hspace{0.5cm}&+& \underbrace{ {\rm d}J(t), }_{\text{Compound jump process}}\\
           &=& \langle u_\text{S}\rangle {\rm d}t \hspace{0.5cm}&+& \sigma {\rm d}W(t) \hspace{0.5cm}&+& {\rm d}J(\Lambda(\epsilon),G(s;\epsilon);t),
\end{align}
where $b(X(t))=\langle u_\text{S}\rangle$ is the mean Stokes drift of a stochastic or irregular non-breaking wave field (angular brackets denote the mean of a stochastic process);  $\sigma (X(t),t) = \sigma $ the standard deviation of the wave-averaged drift rate caused by stochastic nature of the individual waves, with $D=\sigma^2/2$ the resulting  diffusion coefficient; and $W(t)$ denotes a Wiener process (Brownian motion). The effect of wave breaking is captured by the compound Poisson process  $J(t)$, which represents the jumps in particle position induced by breaking. Note this process has a non-zero mean, $\langle u_\text{B}\rangle\neq 0$, reflecting the contribution of breaking to mean drift.

Taking the terms in \eqref{eq:JumpDiffSDE} in turn, we will proceed to outline how particle displacement can be viewed as a drift-diffusion process in the absence of breaking waves (\S \ref{sec:stochastic_Stokes_drift}) and then introduce a compound-jump process to account for wave breaking (\S \ref{sec:breaking_enounters}). Finally, in \S \ref{sec:Fokker_Planck}, we will propose a Fokker--Planck equation for the evolution of the probability density function of particle position $P(X,t)$ and give analytical solutions for the first three moments of $P(X,t)$ (given a Dirac delta function for the initial particle distribution).

\subsection{Stochastic Stokes drift in the absence of breaking}
\label{sec:stochastic_Stokes_drift}
The first term in \eqref{eq:JumpDiffSDE} pertains to the average drift experienced by a particle in the absence of breaking, known as the Stokes drift. For a deep-water, monochromatic, unidirectional wave, the Stokes drift is given by \citep{stokes1847}:
\be\label{eq:stokes:1comp}
U_\text{S}(z) =  a^2 \omega k e^{2k z},
\ee
where $\omega$ is the angular frequency of the wave, $k$ its wavenumber, $a$ its amplitude, and $z$ the vertical position measured upwards from the still-water level. The value of the Stokes drift at the surface $u_\text{S}$ is consistently approximated as $u_\text{S}=U_\text{S}(z=0)$. Written in terms of the (commonly estimated) wave period $T$ and wave height for periodic linear waves $H=2a$:
\be\label{eq:def:StokesSurface}
u_\text{S}  = (ak)^2 c =    \frac{a^2 \omega^3}{g}
= \frac{(2 \pi)^3}{g}\frac{H^2}{T^3},
\ee
where $c=\omega/k$, and we have used the linear deep-water dispersion relationship $\omega^2=gk$ with $g$ the gravitational constant. The Stokes drift is proportional the square of the steepness $ak$, as is evident from \eqref{eq:def:StokesSurface}.

An irregular or stochastic wave field consists of a distribution of different wave periods and heights. The mean Stokes drift can be obtained by summing up the Stokes drift contributions of the different spectral components \citep{Kenyon1969, Webb2011}:
\be\label{eq:stokes:spectr}
\langle u_\text{S} \rangle = \frac{16\pi^3}{g}  \int_0^\infty S(f) f^3   {\rm d}f,
\ee
where the unidirectional frequency spectral density $S(f)$ is defined so that $\int_0^\infty S(f) {\rm d}f =  \langle \eta^2\rangle$, with $\eta(t)$ the surface elevation time series, and $f$ the wave frequency. Naturally, for a monochromatic wave, equation \eqref{eq:stokes:spectr} gives the same result as \eqref{eq:def:StokesSurface}.

The second term in \eqref{eq:JumpDiffSDE}, ${\rm d}X = \sigma  {\rm d}W$, models stochastic deviations from the mean as a Wiener or normal diffusion process. Conceptually, deviations from the mean arise because, on the wave-averaged time scale, each different wavelength and wave height in an irregular sea contributes a different Stokes drift and, therefore, a different displacement. Regardless of the distribution of the Stokes drift, as long as its variance is finite, the central limit theorem states that this will result in a normal distribution for particle position $X(t)$. For a normal process, the second central moment or variance $\langle (X(t)-\langle X(t) \rangle)^2 =2Dt$, where $D$ is the diffusion coefficient. Assuming a stationary underlying random process, the statistics of fluctuations in the drift velocity (i.e., $ \tilde{u}_\text{S} = u_\text{S}-\langle u_\text{S}\rangle$) do not depend on time. One can write  $D=\langle \tilde{u}_\text{S}^2 \rangle \tau = \sigma_\text{S}^2 \tau $ , with $\tau \propto \Delta\omega ^{-1}$ the integral correlation time of the drift velocity, $\Delta \omega$ a measure of the width of the underlying spectrum, and with $\sigma_\text{S}^2=\langle \tilde{u}_\text{S}^2  \rangle$ the variance of the Stokes drift \citep{Taylor1922,Herterich1982,Farazmand2019}. Therefore, $\sigma= \sqrt{2D}=\sqrt{2 \tau}\sigma_\text{S}$ in \eqref{eq:JumpDiffSDE}. 


To estimate the diffusion coefficient $D$, it is therefore necessary to obtain a distribution for the Stokes drift $P(u_\text{S})$ and derive from this its standard deviation $\sigma_\text{S}$. In \cite{Herterich1982} the spectrum is assumed to be narrow, implying a constant (non-stochastic) value for the period $T=T_p$. Consequently, the distribution for the Stokes drift can be derived directly from the Rayleigh distribution for the wave height using \eqref{eq:def:StokesSurface} \citep{Longuet-Higgins1952}. Since $u_\text{S} \propto H_s^2$, the Stokes drift follows an exponential distribution. Specifically, the probability density function for $u_\text{S}$ reads:
\begin{equation}\label{eq:dist:usNB}
    P(u_{\text{S}};H_\text{s}, T_p) =  \frac{g T_p^3}{4 \pi^3}  \frac{1}{H_\text{s}^2} \exp{\left[-\frac{gT_p^3}{ 4\pi^3}\frac{u_\text{S}}{H_\text{s}^2}\right]}.
\end{equation}
Alternatively, \citet{Myrhaug2014} and \citet{Longuet-Higgins1983} derive an exponential distribution for the Pierson--Moskowitz spectrum, taking into account the joint distribution of wave heights and wave periods. In both cases, an exponential distribution is obtained, which has the property that the mean is equal to the standard deviation. We therefore have  $\sigma_\text{S}=\langle u_\text{S}\rangle$, which we will use to predict the diffusion coefficient $D$.

\subsection{Breaking encounters as a jump process}
\label{sec:breaking_enounters}

Our experiments will show (cf. \S \ref{sec:experiment} and 
figure \ref{fig:exampleTraj} in particular) that for waves that are steep enough so that they start to break, in addition to the slow gradual drift and diffusion of the particles predicted for non-breaking waves, jumps in particle positions occur when waves break. Each `jump' event represents an encounter of a particle with (the crest of) a breaking wave. We model these jump events by a compound Poisson process $J(t)$ in \eqref{eq:JumpDiffSDE}:
\begin{equation}
    J(t) = \sum_{i=0}^{N_\Lambda(t)}s_i.
\end{equation}
where $N_\Lambda(t)$ is a counting of a Poisson point process, and the arrival rate $\Lambda$  corresponds to the expected number of jumps per unit time for the compound Poisson process. 

We expect the arrival rate $\Lambda$ to increase with the steepness $\eps$. Assuming a gradual increase  of the arrival rate with steepness followed by followed by saturation at large enough steepness, we propose a three-parameter sigmoid function (see figure \ref{fig:lambda_G}a): 
\begin{equation}\label{eq:lambda}
    \Lambda(\eps; \tau_\Lambda, \phi_\Lambda, \eps_{0,\Lambda}) = \frac{1/ \tau_\Lambda}{1+\exp{[-\phi_\Lambda(\eps-\eps_{0,\Lambda})}]},
\end{equation}
where the parameters $\tau_\Lambda$, $\phi_\Lambda$ and $\eps_{0,\Lambda}$ will be estimated from our experimental data (see \S \ref{sec:experiment}). Furthermore, we assume that the amplitude of the jumps $s_i$ follows a two-parameter Gamma distribution with probability density function $G(s;\alpha(\epsilon),\beta(\epsilon))$ (see figure \ref{fig:lambda_G}b):
\begin{equation}\label{eq:gammaDistr}
G(s;\alpha,\beta) = \frac{\beta^\alpha}{\Gamma(\alpha)}s^{\alpha-1}e^{-\beta s},
\end{equation}
where $\alpha>0$ is the shape parameter, $\beta>0$ the rate parameter and $\Gamma(\alpha)$ a Gamma function. Based on experimental data, we will show in \S \ref{sec:experiment} that both $\alpha$ and $\beta$ can be effectively modelled as linear functions of steepness $\epsilon$.

\subsection{Fokker--Planck equation and analytical solutions}
\label{sec:Fokker_Planck}
Two methods exist to integrate the stochastic differential equation \eqref{eq:JumpDiffSDE} in order to obtain a probability distribution for particle position $X(t)$. Monte Carlo simulations of \eqref{eq:JumpDiffSDE} can be numerically integrated (e.g., \citet{milstein_numerical_1995,Higham2001}), and an empirical probability density function can be obtained at each time step from the statistics of these trajectories. Alternatively, the evolution equation for the probability density function $P(x,t)$ of the random variable $X(t)$ corresponding to \eqref{eq:JumpDiffSDE}, the so-called Fokker--Planck equation, can be solved directly. For \eqref{eq:JumpDiffSDE}, the Fokker--Planck equation is given by (e.g., \cite{gardiner_handbook_1983,Denisov2009,gaviraghi_analysis_2017}):
\begin{equation}\label{eq:FPjumpdiff}
    \frac{\partial}{\partial t}P(x,t) = -\langle u_\text{S} \rangle \frac{\partial}{\partial x}P(x,t) + \frac{\sigma^2}{2} \frac{\partial^2}{\partial x^2}P(x,t) - \Lambda P(x,t) + \Lambda \int_{-\infty}^{\infty}{\rm d}x' P(x',t) J(x-x').
\end{equation}
This partial differential equation can be solved numerically (e.g., \cite{Gaviraghi2017}). Alternatively, by employing the characteristic functions of the distribution, 
\begin{equation}
    \hat{P}(l) = \int_{-\infty}^{\infty}{\rm d}x e^{ilx} P(x),
\end{equation}
we obtain the ordinary differential equation
\begin{equation}\label{eq:FPjumpdiffFT}
   \frac{\partial}{\partial t} \hat{P}(l)  = i \langle u_\text{S} \rangle l  \hat{P}(l) - \frac{\sigma^2}{2}l^2 \hat{P}(l)  - \Lambda \hat{P}(l)  + \Lambda \hat{G} \hat{P}(l), 
\end{equation}
where the characteristic function of the gamma distribution is
\begin{equation}
    \hat{G}(l) = \left(1-i \frac{l}{\beta}\right)^{-\alpha}.
\end{equation}
\Cref{eq:FPjumpdiffFT} can be solved exactly, with solution:
\begin{equation}
    \hat{P}(l,t) = A(l) \exp{\left[\left(i \langle u_\text{S} \rangle l - \frac{\sigma^2}{2}l^2 - \Lambda + \Lambda \left(1-i \frac{l}{\beta}\right)^{-\alpha}\right)t\right]},
    \label{eq:soln_general}
\end{equation}
where $A(l)$ is an unknown function that is determined by the initial condition. 

Raw moments of a probability density function can be readily evaluated using its characteristic function:
\begin{equation}
    m_n = \langle X^n\rangle =  \int_{-\infty}^{\infty}{\rm d}x x^n e^{ilx} P(x) =\left. (-i)^n \frac{\partial^n \hat{P}(l)}{\partial dl^n}\right\vert_{l=0}.
\end{equation}
For a Dirac delta function as the initial condition (i.e., $\hat{P}(t=0)=1$), we have $A=1$ in \eqref{eq:soln_general} and we can obtain exact solutions for the first (three) central moments of \eqref{eq:FPjumpdiff}:
\begin{alignat}{3}
\langle X(t) \rangle &=& \left.-i \frac{\partial \hat{P}(l,t)}{\partial l} \right\vert_{l=0}& &=&  \left(\langle u_\text{S} \rangle  + \frac{\alpha}{\beta}\Lambda\right)t, \label{eq:sol_m1}\\
         \langle \tilde{X}(t)^2 \rangle &=& \left.- \frac{\partial^2 \hat{P}(l,t)}{\partial l^2}\right\vert_{l=0}  - m_1^2& &=& \left(\sigma^2 + \frac{\alpha (\alpha +1)}{\beta}\Lambda \right) t, \label{eq:sol_m2}\\
    \langle \tilde{X}(t)^3 \rangle &=& \left.i \frac{\partial^3 \hat{P}(l,t)}{\partial l^3} \right\vert_{l=0}  - 3 m_1 m_2 + 2 m_1^3& &=& \frac{\alpha(1+\alpha)(2+\alpha)}{\beta^2}\Lambda t,\label{eq:sol_m3}
\end{alignat}
where $\tilde{X}(t)=X(t)-\langle X(t) \rangle$. In \eqref{eq:sol_m1}-\eqref{eq:sol_m3}, the expected Stokes drift for non-breaking waves $\langle u_\text{S} \rangle (\epsilon)$, the arrival rate $\Lambda(\eps)$, and the shape and scale factors $\alpha(\eps)$ and $\beta(\eps)$ are all functions of steepness; their dependence on $\epsilon$ will be estimated from experimental data in the next section. Note from \eqref{eq:sol_m1}-\eqref{eq:sol_m3} that each central moment is linearly increasing with time. Breaking increases both the mean drift (cf. \eqref{eq:sol_m1}) and the variance of particle position (cf. \eqref{eq:sol_m2}); it also introduces a non-zero (positive) skewness not predicted for non-breaking waves (cf. \eqref{eq:sol_m3}).

\section{Laboratory experiments} \label{sec:experiment}

\subsection{Laboratory set-up and input conditions}

\begin{figure}
  \centerline{\includegraphics[width=\textwidth]{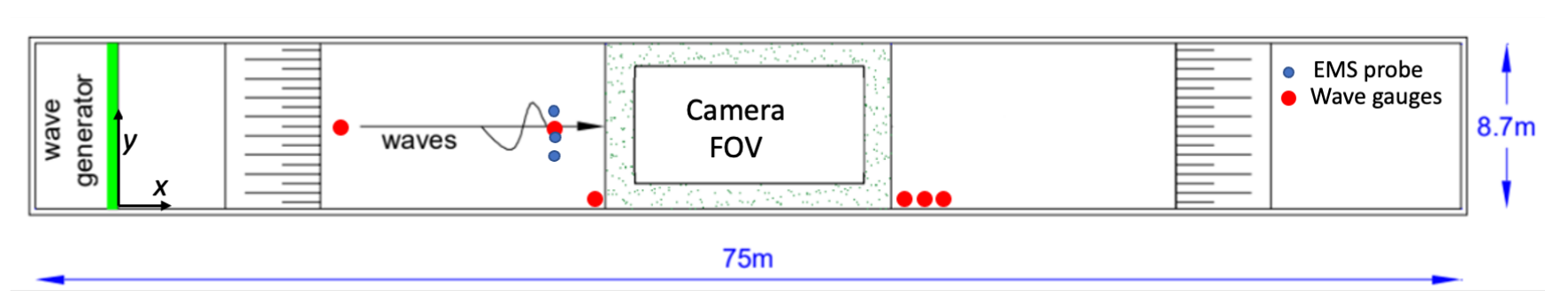}}
  \caption{Experimental set-up, indicating the overhead particle-tracking camera's field of view (FOV), wave gauges (red dots), and EMS velocity measurement probes (blue dots).}
\label{fig:expSetup}
\end{figure}

\subsubsection{Wave conditions}
Experiments were performed in the  8.7 m wide and 75 m long Atlantic Basin at Deltares, the Netherlands. \Cref{fig:expSetup} provides an overview of the set-up. The basin was equipped with  a segmented piston-type  wavemaker, consisting  of  20 wave paddles at one end and an absorbing beach at the other. The water depth $d$ was 1 m. The experiments were carried out with irregular waves prescribed by the JONSWAP spectral density $S(\omega)$,
\be \label{eq:JONSWAP}
S(\omega) = \frac{K g^2}{\omega^5}\exp\left[-\frac{5}{4}\left(\frac{\omega_p}{\omega}\right)^4\right]\gamma^r,
\ee
with $\omega$ the frequency of the waves (in rad/s), $g$ the gravitational acceleration, $\omega_p$ the peak frequency, and $r=\exp\left[-(\omega-\omega_p)^2/(2\sigma_\text{S}^2\omega_p^2)\right]$ with (non-dimensional) spectral width $\sigma_\text{S}= 0.07$ for $\omega \leq \omega_p$ and $\sigma_\text{S} = 0.09$ for $\omega>\omega_p$. The shape parameter $\gamma$ captures the `peakedness' and thereby also the bandwidth of the spectrum, and we set $\gamma=3.3$ for all experiments. The bandwidth is important as it determines the correlation time $\tau$ and thereby the variance of particle position predicted by \eqref{eq:JumpDiffSDE} (i.e., $\sigma= \sqrt{2D}=\sqrt{2 \tau}\sigma_\text{S}$). Using a second moment of the spectrum, we calculate the spectral width as $\Delta \omega
= 1.39 $ rad/s and set $\tau=1/\Delta \omega$. The prefactor $K$ in \eqref{eq:JONSWAP} is adjusted to obtain the desired significant wave height $H_s$. Phases of a discretised spectrum were chosen randomly in order to create an irregular wave times series of 30 min duration, which was used as (linear) forcing to the wavemakers. Reflections were generally less than 5\%.

\begin{table}
  \begin{center}
\def~{\hphantom{0}}
  \begin{tabular}{lcccccccccc}
     $H_s$ & $H_\text{s,exp}$ &$\eps_\text{exp}$  &
     $\Delta t$& $N_\text{traj}$ & 
     $\Lambda$ & $\langle u_{\text{S}} \rangle$  & $\langle u_\text{L,exp} \rangle$ & $\langle u_\text{E,fit} \rangle$  \\[3pt]
    (m) & (m) &    & (s)   &   & (1/s)  & (mm/s)  &  (mm/s) & (mm/s) \\[3pt]
    0.050 & 0.053 &0.074 & 257 & 143 &  $3.25\times 10^{-5}$ &   13.1     &     13.7  & +4.1 \\
    0.090 & 0.087 &0.122  & 167 & 136 & $1.42\times 10^{-2}$ & 24.9     &     22.5  & +0.8  \\
    0.120 & 0.115 &0.162 & 122 & 85 &  $6.96\times 10^{-2}$  & 35.3     &     30.0  & -5.7\\
    0.170 & 0.132 &0.185 & 143 & 113 & $6.59\times 10^{-2}$ & 43.8     &     31.2  & -10.8 \\
  \end{tabular}
  \caption{Overview of experiments and parameter values. For all experiments $T_p = 1.2$ s. The subscript exp refers to the values measured in experiments, $\Delta t$ is the time length of a (segmented) trajectory, $N_\text{traj}$ corresponds to the number of spheres tracked, and $\Lambda$ is the arrival rate of jumps in particle position. The mean Stokes drift $\langle u_{\rm S}\rangle$ is calculated using \eqref{eq:stokes:spectr}, the mean Lagrangian velocity $\langle u_{\rm L,exp}\rangle$ is obtained from tracer particle positions in experiments, and the Eulerian-mean velocity $\langle u_{\rm E,fit}\rangle$ is obtained so that experiments and model predictions for the mean agree (see \S \ref{sec:Eulerian_mean_flow}).}
  \label{tab:experiment}
  \end{center}
\end{table}

Parameter values chosen for the experiments are given in table \ref{tab:experiment}. The peak period $T_p=1.2$ s, and four (input) significant wave heights are examined, $H_s =$ 0.05, 0.09, 0.12, and 0.17 m (input), with an increasing fraction of breaking waves. Defining a characteristic steepness as $\eps =k_p H_s/2$, with $k_p$ the wavenumber corresponding to $T_p$, we obtain $\eps =$ 0.070, 0.126, 0.168, and 0.238 (input). Experiments were performed in deep water ($k_p d=2.8$). A total of 
6 wave gauges were placed in the basin. The time series of the surface elevation $\eta(t)$ at wave gauge 2 (co-located with the EMS probes) was used to calculate the
values of the parameters measured in experiments reported in table \ref{tab:experiment}, where they are labelled with the subscript expt to be distinguished from input values, and the power spectrum $S(f)$ (used to estimate the Stokes drift according to  \eqref{eq:stokes:spectr}). From table 
\ref{tab:experiment} 
it is evident that
$H_s$ is under-produced in the experiments for the larger values of $H_s$, which is in large part due to wave breaking. 

\subsubsection{Lagrangian tracer particles}
The tracer particles were 20 mm diameter yellow polypropylene spheres, which were chosen to be as small as possible, but large enough to be tracked by the overhead camera. The density of the particles was 920 kg/m$^3$, so the particles were as submerged as possible so that they would follow the motion of the fluid, but remained detectable from above. The experiments were carried out in fresh water (998 kg/m$^3$). An automated device was used to drop a set of approximately 20 spheres into the basin every 10 s, with a spacing of 15 cm along a 3.0 m spanwise section of the basin ($y$-direction) a short distance (in the $x$-direction) before entering the camera's field of view. A Z-CAM camera was mounted at a height of 11.9 m above the basin, allowing it to capture an area of approximately 8 m along the length of the basin and 6 m of its width (see figure \ref{fig:expSetup}).

To examine whether our tracer particles behave as idealized Lagrangian particles, we compare the measured velocity spectrum  of the tracer particles to the velocity spectrum obtained from the measured surface elevation using linear wave theory in \cref{app:velocityspec}. For frequencies up to a high-frequency cut-off that lies much above the spectral peak, we find good agreement between the two, confirming Lagrangian behaviour of the tracer particles. Based on  \cite{Calvert2021}, we estimate that spherical particles with a diameter of less than 10\% of the wavelength typically behave as Lagrangian particles. For $20$ mm diameter spheres, this corresponds to a cut-off frequency of $2.8$ Hz (from linear dispersion), which in turn agrees with what we find in the aforementioned spectral comparison in \cref{app:velocityspec}. Since $D/\lambda_p=0.9\%$, we do not expect our tracer particles to experience enhanced transport compared to the Stokes drift in non-breaking waves due to mechanisms described in \cite{Calvert2021}. 

\subsubsection{Role of Eulerian-mean flow}
\label{sec:Eulerian_mean_flow}
While we do not take into account the Eulerian-mean flow $u_\text{E}$ in our model \eqref{eq:JumpDiffSDE}, this flow is present in experiments and therefore has to be accounted for to enable comparison (cf. \eqref{eq:Lagrangian-mean}). Wave-induced Eulerian-mean flows have often prevented observation of Stokes drift in laboratory wave flumes; they are notoriously difficult to predict, specific to each laboratory basin and experiment and, when observed in the laboratory, not representative of wave-induced Eulerian-mean flows in the field (see reviews by \citet{vandenbremer_breivik2017} and \cite{monismith2020}). 

The Eulerian flow was measured using three EMS velocity measurement devices at three different depths (always fully submerged) and one horizontal position (($x,y=14.2, 4.05$-$4.65)$ m). This allowed estimation of the basin-specific  Eulerian-mean flow for each experimental condition at fixed depths of $z=$ $-20$, $-30$, and $-50$ cm with $z=0$ the still-water level.

Although we have measured the Eulerian flow directly, we do not use the Eulerian-mean flow that we obtain from these measurements (by wave averaging) directly in the comparison between experiments and model predictions (cf. \eqref{eq:Lagrangian-mean}). The reason is that these Eulerian flow measurements, for reasons of practicality, are only at a single point in space ($x$, $y$) and at a certain distance below the free surface, making extrapolation to the surface a potential source of error. Nevertheless, we can infer from the measurements that the Eulerian-mean flow is non-stochastic, that is, it does not show variability on the same time scale and with the same order of magnitude as the measured Lagrangian-mean velocity. 

We therefore correct $b$, the mean non-breaking drift in our model \eqref{eq:JumpDiffSDE}, with an arbitrary (not measured) Eulerian-mean flow 
$\langle u_{\rm E, fit}\rangle$, so that $b=\langle u_{\rm S}\rangle+\langle u_{\rm E, fit}\rangle$. The mean Stokes drift $\langle u_{\rm S}\rangle$ is calculated using \eqref{eq:stokes:spectr} from (a JONSWAP spectrum fitted to) the measured surface elevation spectrum. The arbitrary value of the Eulerian-mean flow 
$\langle u_{\rm E, fit}\rangle$ is then chosen so that the mean Lagrangian drift predicted by the model $\langle u_\text{L,mod} \rangle ={\rm d}\langle X(t)\rangle/{\rm d}t $ (using $b=\langle u_{\rm S}\rangle+\langle u_{\rm E, fit}\rangle$) is equal to the mean Lagrangian drift in experiments $\langle u_\text{L,exp}\rangle$, that is $\langle u_\text{L,mod} \rangle=\langle u_\text{L,exp} \rangle$. Our interest in \S \ref{sec:results}, where we compare model predictions to experiments,
is therefore in higher-order moments of particle position. The values $\langle u_{\rm E, fit}\rangle$ thus obtained are within a reasonable margin of the values measured at depth $\langle u_{\rm E, exp}\rangle$ (see \cref{tab:uLdifference} in \cref{app:Velocities}). We note for completeness that for non-breaking waves, we expect that $\langle u_\text{L,mod} \rangle =  \langle u_\text{S} \rangle + \langle u_\text{E,fit} \rangle$, while for breaking waves the jumps also make a contribution to the mean, $\langle u_\text{B}\rangle$. The next section will explain how this contribution is calculated, which must be done before  $\langle u_{\rm E, fit}\rangle$ can be found. Table \ref{tab:experiment} gives the values of the different mean velocities for the different experiments.

\subsection{Trajectory data processing}
Our goal is to predict particle trajectories and properties of the probability distribution of particle position based on information of the spectrum of the waves. To this end, we first have to process the camera images to obtain particle trajectories. We subsequently use these trajectories to obtain properties of the jump process that is used to model breaking.

\subsubsection{Image processing}
The yellow spheres were tracked using OpenCV. The spheres  were identified using a Hue saturation filter and subsequently tracked using the CSRT algorithm, obtaining the sub-pixel location ($x_p,y_p$) of the centre of each sphere at each point in time. The time step is determined by the frame rate of the camera of 24 Hz. The trajectories were undistorted by calculating the camera intrinsics using a checkerboard with 75 mm squares. The pixel locations were then transformed into basin coordinates ($x,y$), assumed to be in the plane of the still-water level, $z=0$, defined by a floating checkerboard at a known position. See appendix \ref{app:DataProcessing} for further details.

\subsubsection{Creating sample trajectories}
\begin{figure}
  \centerline{\includegraphics[width=135mm]{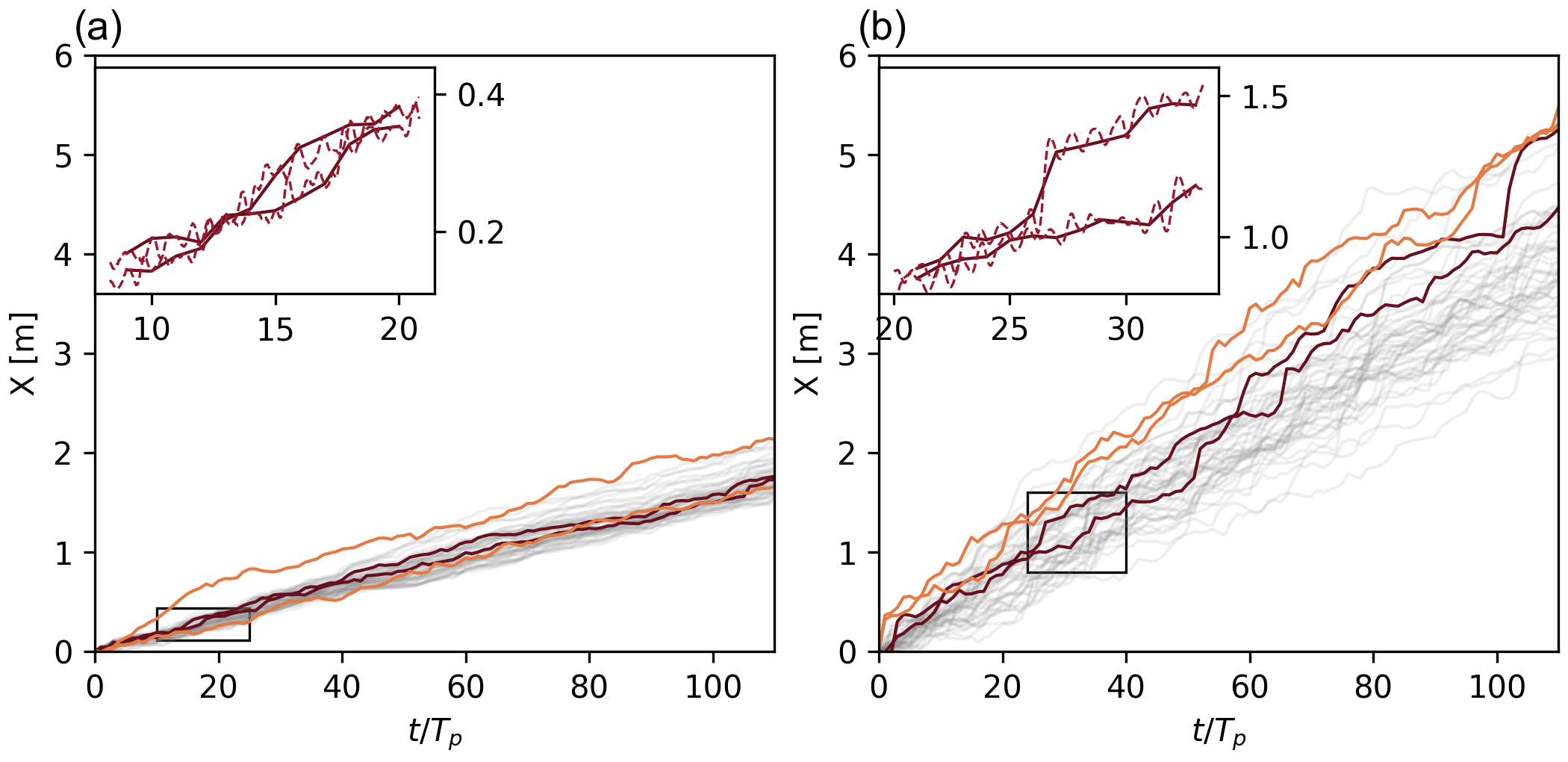}}
  \caption{Example particle trajectories in irregular waves, showing experiments (light gray, and dark red) and Monte Carlo simulations of our model (orange): 
  (\textit{a}) $H_s$ = 0.05 m, showing almost no breaking events and trajectories evolving according to a Brownian motion, (\textit{b}) $H_s$ = 0.17 m, showing particles regularly surfing on a breaking wave, causing a jump in the position. This is modelled by discrete jumps in the simulations. The insets show a zoomed-in view of the dark red lines, where the red dashed lines show an oscillating trajectory, corresponding to the effect of waves, corresponding to the experimental data before wave averaging. All the other (solid) lines show the wave-averaged position data (obtained from averaging using a period corresponding to the peak period of the waves $T_p$).}
\label{fig:exampleTraj}
\end{figure}

To create samples that can be used to compare to predictions of our stochastic model, particle trajectories $X_i(t)$ were segmented to trajectory lengths of $\Delta t$ and all offset to have initial position $X_i(t=0) = 0$, as illustrated in figure \ref{fig:exampleTraj}. The resulting initial distribution is a Dirac delta function, and we obtain $N_{\rm traj}$ trajectories of equal length $\Delta t$ (for each significant wave height).

The dashed lines in the insets in figure \ref{fig:exampleTraj} show particle position $X(t)$ at the time resolution of the camera (24 Hz), showing the oscillatory motion  of the particle with every wave. The stochastic model  \eqref{eq:JumpDiffSDE} is valid for a large enough time scale, so that the oscillatory effects of the waves are averaged out, but the stochastic effect of the irregular wave field on particle transport is retained. We therefore sample the trajectories at  time interval $T_p$ to obtain stochastic wave-averaged trajectories. These are indicated by the solid lines in the insets in figure \ref{fig:exampleTraj} (and by all the lines in figure \ref{fig:exampleTraj} itself). 

\subsubsection{Breaking detection}
For the lowest significant wave height we have considered, breaking is negligible, whereas for the highest, the particles have many encounters with breaking crests, as indicated in \cref{tab:experiment} by the jump arrival rate $\Lambda$, which measures the number of jumps per unit time. \Cref{fig:exampleTraj} illustrates the evolution of particle position for both these extremes. 

We classify particle motion as a `jump', when the instantaneous horizontal velocity (obtained from the trajectories before wave averaging) surpasses a velocity threshold set as $u_{th} = 0.3c$, where $c=\omega/k$ is the phase velocity obtained from the linear dispersion relationship. Although this threshold is arbitrary, lowering it results in a high number of jump events for $H_s=0.05$ m, whilst breaking only rarely occurred for these experiments. Physically, the threshold reflects the idea that particles during breaking are transported with the crest at the phase velocity of the wave $c$ (they `surf' the wave) instead of the much smaller Stokes drift velocity. The distance covered at velocities higher than this threshold during one breaking event is the jump amplitude $s$. See \cref{app:JumpDetection} for an example of this breaking detection procedure.

\begin{figure}
  \centerline{\includegraphics[width=140mm]{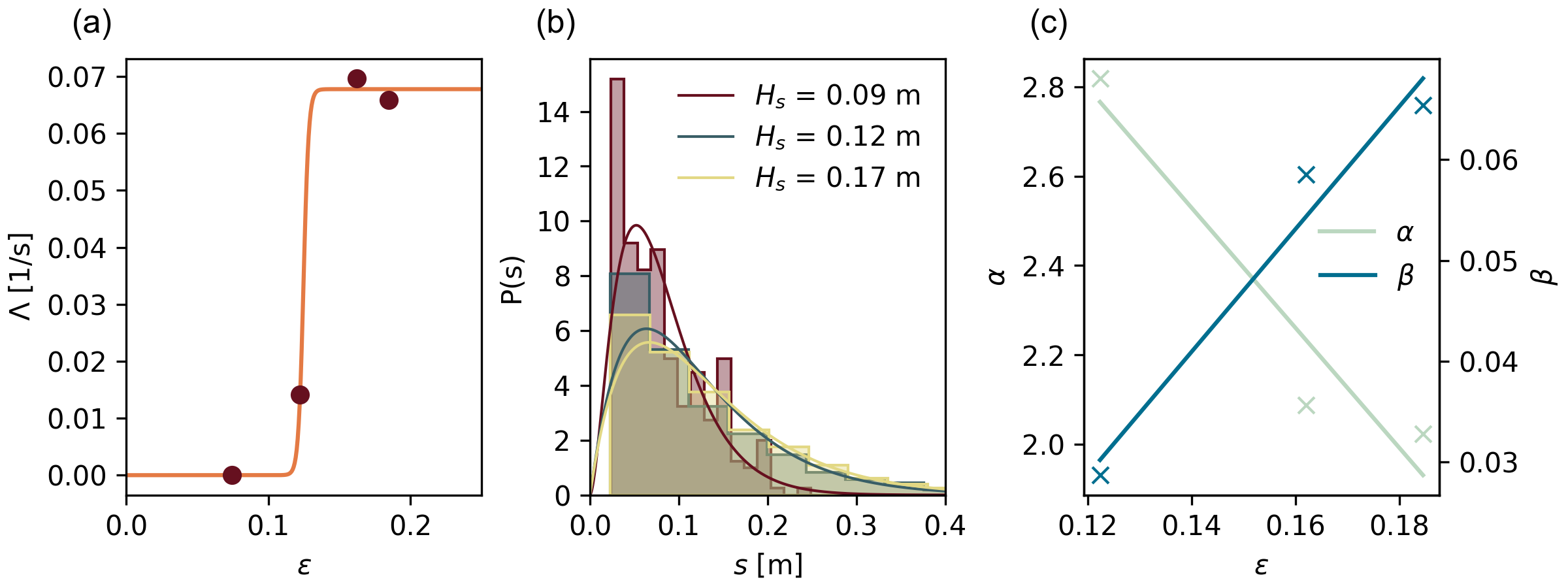}}
  \caption{Calibration of the jump process for transport by breaking waves: (\textit{a}) jump arrival rate $\Lambda$ as a function of steepness $\eps$, (\textit{b}) probability density function of jump size $G(s) = \Gamma(s;\alpha,\beta)$ for $H_s$ = 0.09 (red), 0.12 (green) and 0.17 m (yellow) (the case of $H_s$ = 0.05 m is not shown, as no jumps are detected for this case), (\textit{c}) scale and shape parameters $\alpha$ and $\beta$ for the probability density function of jump size as a function of $\eps$ as given by \eqref{eq:scale_and_shape_parameter}.}
\label{fig:lambda_G}
\end{figure}

We use the jumps thus obtained to calibrate the jump process described by \eqref{eq:lambda}-\eqref{eq:gammaDistr}. Figure \ref{fig:lambda_G}a shows the jump arrival rate $\Lambda$ as a function of steepness $\epsilon$ estimated from the experimental data for the four values of steepness. Also shown is the sigmoid function for $\Lambda(\epsilon)$ given by \eqref{eq:lambda} with estimated values of the parameters in table \ref{tab:breaking}. The variation in jump amplitudes $s$ estimated from the experimental data in figure \ref{fig:lambda_G}b is captured well by the Gamma distribution \eqref{eq:gammaDistr}. Finally, we estimate the parameters of the Gamma distribution \eqref{eq:gammaDistr} as linear function of steepness:
\begin{equation}
    \alpha=a_\alpha+b_\alpha \epsilon, \quad
    \beta=a_\beta+b_\beta \epsilon,
    \specialnumber{a,b}
    \label{eq:scale_and_shape_parameter}
\end{equation}
as shown in figure \ref{fig:lambda_G}c with coefficients in table \ref{tab:breaking}. 

\begin{table}
  \begin{center}
\def~{\hphantom{0}}
  \begin{tabular}{ccccccc}

    \multicolumn{3}{c}{$\Lambda(\eps)$} & \multicolumn{4}{c}{$G(s;\alpha(\eps),\beta(\eps))$}\\
\cmidrule(r){1-3} \cmidrule(r){4-7}
    $\tau_\Lambda$ [s] & $\phi_\Lambda$  & $\eps_{0,\Lambda}$ &$a_\alpha$ & $b_\alpha$ & $a_\beta$  & $b_\beta$   \\
    \midrule 
     0.0677   &503  &   0.128  & -6.49  & 3.46, & 0.305 & -0.003 \\
  \end{tabular}
  \caption{Calibrated values of the parameters describing the dependence of the jump process for breaking waves on steepness $\eps$, distinguishing the jump arrival rate $\Lambda(\epsilon)$ according to \eqref{eq:lambda} and the jump size probability density function $G(s)$ according to \eqref{eq:gammaDistr}.}
  \label{tab:breaking}
  \end{center}
\end{table}

\section{Results}\label{sec:results}
In this section, we will compare Monte Carlo simulations of our model \eqref{eq:JumpDiffSDE} (using $10^5$ trajectories) to experiments. The Monte Carlo simulations agree perfectly with the exact solutions for the first three moments \eqref{eq:sol_m1}-\eqref{eq:sol_m3}, so we will only show the former. We will examine non-breaking (\S \ref{sec:Non-breaking_waves}) and breaking waves (\S \ref{sec:Breaking_waves}) in turn, followed by model predictions as a function of steepness (\S \ref{sec:Model_predictions}).

\subsection{Non-breaking waves: comparison between experiments and model predictions}
\label{sec:Non-breaking_waves}
\begin{figure}
  \centerline{\includegraphics[width=150mm]{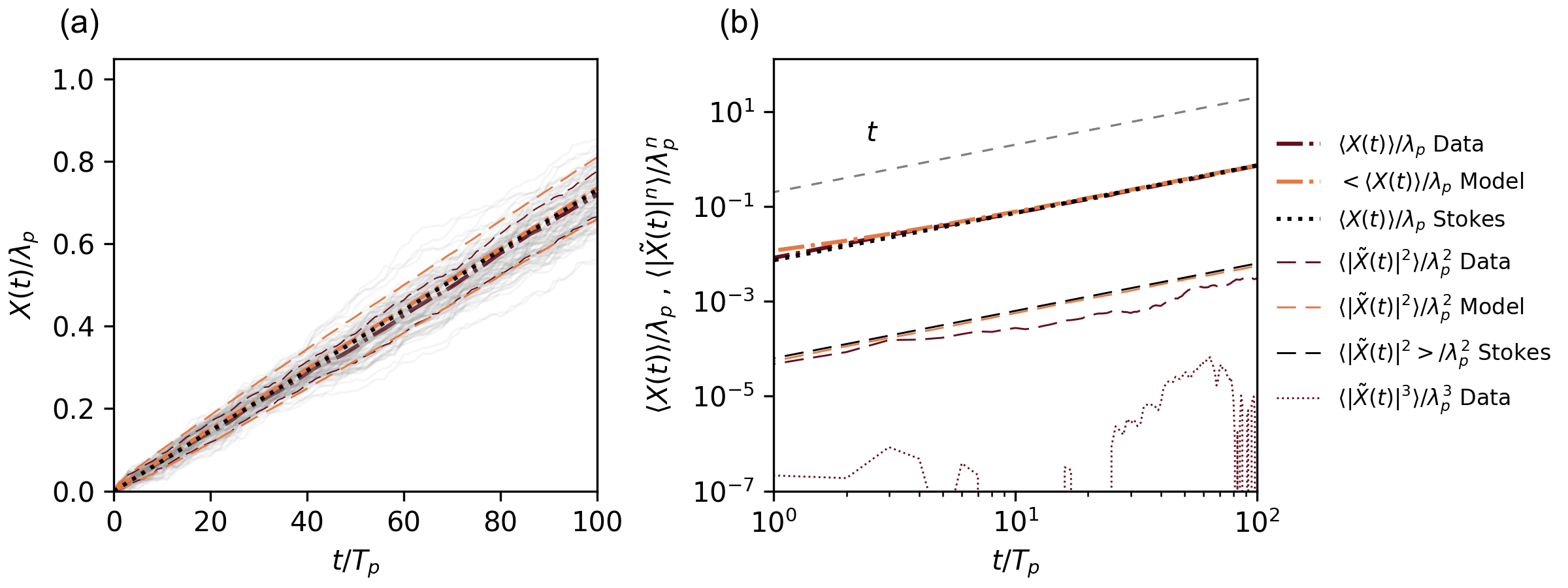}}
  \caption{Comparison between model and experimental data for non-breaking irregular waves ($H_s$ = 0.05 m): (\textit{a}) normalized mean position $\langle X(t) \rangle/\lambda_p$ (dashed-dotted) and $\pm 1 $ standard deviation (dashed) for experiments (red) and model simulations (orange), (\textit{b}) normalized mean position (dashed-dotted) for experiments (red), model simulations (orange) and Stokes theory for non-breaking waves (dotted black), showing proportionality to $t$. Normalized variance $\langle|\tilde{X}(t)|^2\rangle/\lambda_p^2$ (dashed) for experiments (red) and model simulations (orange) are approximately proportional to time $t$. The normalized skewness $\langle|\tilde{X}(t)|^3\rangle/\lambda_p^3$ (dotted) is negligible in the experiments }
\label{fig:compHs0p05}
\end{figure}

\Cref{fig:compHs0p05}a shows the normalized mean $ \langle X(t)\rangle/\lambda_p$ (dashed-dotted), where $\lambda_p$ is the peak wavelength, for the experiments (red) and the Monte Carlo simulations of \eqref{eq:JumpDiffSDE} (orange) for the smallest steepness waves ($H_s=$0.05 m, $\eps=0.074$). A negligible number of waves break for this case such that $\Lambda \approx 0$, and the wave breaking term does not contribute in \eqref{eq:JumpDiffSDE}. Therefore, the mean displacement by the Stokes drift (dotted line) coincides with the mean drift predicted by the model, which in turn is equal to that observed in the experiments (by definition here, as we have used this agreement to estimate the Eulerian-mean flow, see \S \ref{sec:Eulerian_mean_flow}). The dashed lines show $\pm 1$ standard deviation.

More importantly, the normalized variance $\langle|\tilde{X}(t)|^2\rangle/\lambda_p^2$, with $\tilde{X}(t)=X(t)-\langle X(t) \rangle$, of the model simulations follows the experiment closely in figure \ref{fig:compHs0p05}b. Indeed, extracting the power-law behavior in figure \ref{fig:compHs0p05}b, the experimental particle position variance exhibits a linear $t$ dependence, in accordance with the single-particle Taylor diffusion prediction by \citet{Herterich1982}, which in turn agrees with our theoretical prediction \eqref{eq:sol_m2} when $\Lambda=0$ (no breaking). It is interesting to note that a Wiener or normal diffusion process has a linear dependence on time, in contrast to either sub- or superdiffusion, for which the conditions of the central limit theorem are violated. Note that this result (the linear dependence on time) is disagreement with  \cite{Farazmand2019}, who predict a superdiffusion $\langle|\tilde{X}(t)|^2\rangle \propto t^4$ for $t > T_p$ based on the  nonlinear John--Sclavounos equation. The particle distribution we observe remains Gaussian with zero skewness (see figure \ref{fig:compHs0p05}), as predicted by the analytical solutions for the third central moment in \eqref{eq:sol_m3}.
 

\subsection{Breaking waves: comparison between experiments and model predictions}
\label{sec:Breaking_waves}
\begin{figure}
  \centerline{\includegraphics[width=150mm]{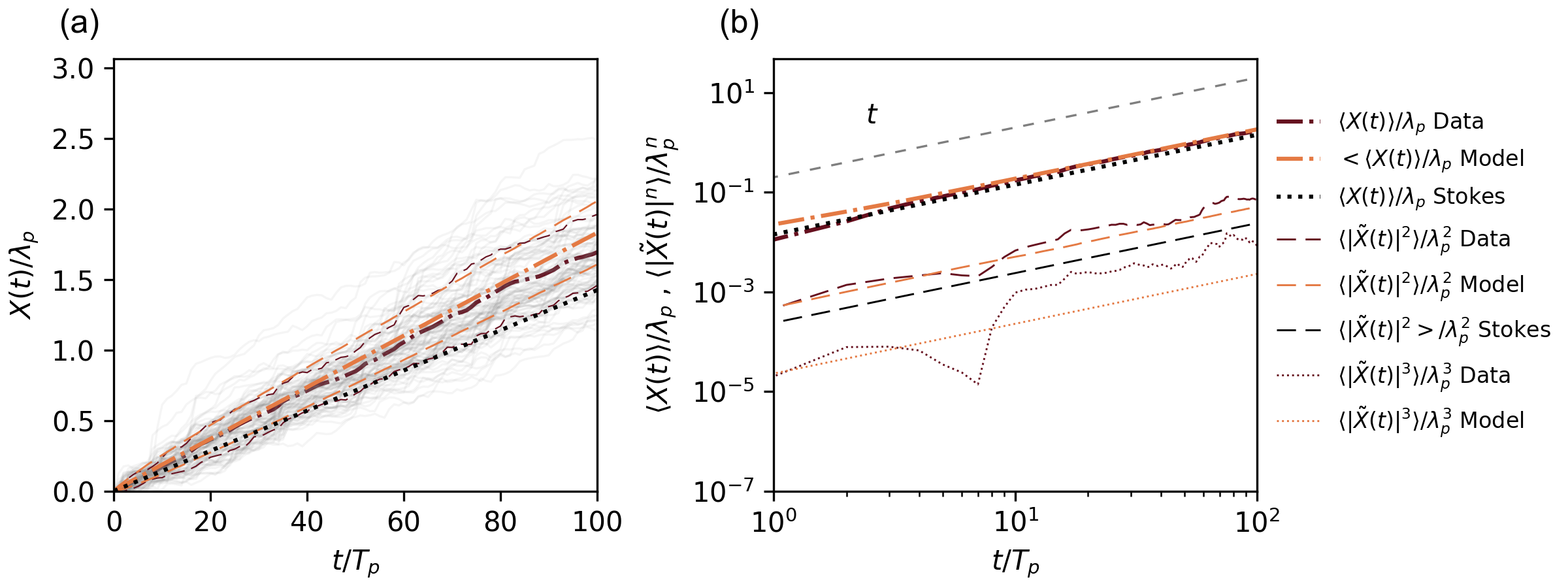}}
  \caption{Comparison between model and experiment data for breaking irregular waves ($H_s$ = 0.17 m): (\textit{a}) normalized mean position $\langle X(t) \rangle/\lambda_p$ (dashed-dotted) and $\pm 1$ standard deviation (dashed lines) for experiments (red) and model simulations (orange), (\textit{b}) normalized mean position (dashed-dotted) for experiment (red), model simulations (orange) and Stokes theory for non-breaking waves (dotted black), showing proportionality to $t$. Normalized variance $\langle|\tilde{X}(t)|^2\rangle/\lambda_p^2$ (dashed) for experiments (red) and model simulations (orange) are approximately proportional to time $t$. The normalized skewness $\langle|\tilde{X}(t)|^3\rangle/\lambda_p^3$ (dotted) is finite for this wave steepness with order-of-magnitude agreement between experiments (red) and model simulations (orange).}
\label{fig:compHs0p17}
\end{figure}

\Cref{fig:compHs0p17}a shows the mean particle position (dashed-dotted) $\pm 1$ standard deviation (dashed lines) in the experiments (red) and the Monte Carlo simulations of \eqref{eq:JumpDiffSDE} (orange) for the largest steepness waves ($H_s=0.17$ m and $\eps=0.185$). For this significant wave height, many jumps in position due to breaking occur, as illustrated earlier in figure \ref{fig:exampleTraj}b. Due to the jump events, the mean drift (the slope of the line in figure \ref{fig:compHs0p17}a) is higher than the theoretical prediction by the Stokes drift (dotted black line) and instead follows that of \eqref{eq:sol_m1} with $\Lambda\neq 0$. The power-law behavior in figure \ref{fig:compHs0p17}b indicates that, in agreement with \eqref{eq:sol_m2}, the variance still scales linearly with time. Here, the diffusion based on the Stokes drift alone (black dashed line) underestimates the measured diffusion. Adding the effect of breaking according to \eqref{eq:sol_m2} gives good agreement with experimental data, demonstrating that the contribution of breaking to particle diffusion can be modelled effectively as a 
Poisson process. Finally, the finite skewness predicted by the model \eqref{eq:sol_m3} is also observed in experiments (orange and red dotted lines in figure \ref{fig:compHs0p17}b, respectively).

\subsection{Model predictions as a function of steepness}
\label{sec:Model_predictions}

\begin{figure}
  \centerline{\includegraphics[width=145mm]{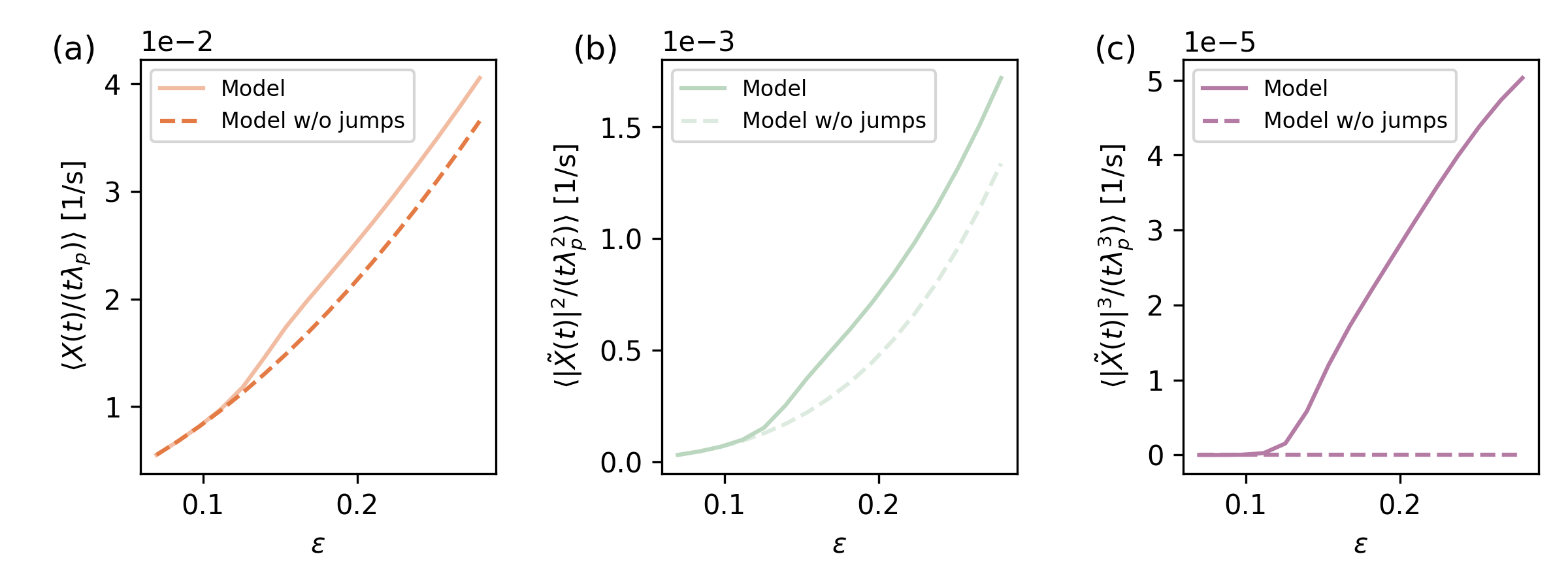}}
  \caption{Model predictions with and without the jumps that transport by breaking waves, showing (\textit{a}) normalized mean velocity as a function of characteristic steepness $\eps$, (\textit{b}) variance of particle position, and (\textit{c}) skewness of particle position.}
\label{fig:simQuant}
\end{figure}

To quantify how the mean velocity and variance and skewness of particle position are affected by the number of breaking encounters, we perform simulations for the characteristic steepness range $\eps = [0.05,0.3]$ setting the Eulerian-mean flow to zero, as summarized in figure \ref{fig:simQuant}. 

\Cref{fig:simQuant}a shows the mean drift $\langle u_{\rm L,mod} \rangle = {\rm d}\langle X(t)\rangle/{\rm d}t$ based on \eqref{eq:JumpDiffSDE}. If the breaking term is not taken into account in \eqref{eq:JumpDiffSDE} (i.e., $\Lambda=0$), this  drift would be equal to the theoretical Stokes drift value $\langle u_\text{S}\rangle \propto \eps^2 $ (dashed line). Taking into account the jump-diffusion process of \eqref{eq:JumpDiffSDE}, $\langle u_{\rm L,mod} \rangle$ starts deviating from  $\langle u_\text{S} \rangle$ for steep waves in which many breaking events occur. The nature of this deviation is given in \eqref{eq:sol_m1}, where $\langle u_{\rm L,mod} \rangle$ is increased by a factor $(1 + \alpha\Lambda/(\langle u_\text{S} \rangle \beta))$. The second moment or variance deviates in a similar fashion from the non-breaking case in figure \ref{fig:simQuant}b, deviating by a factor $\left(1 + \alpha (\alpha +1)\Lambda/(\sigma^2 \beta) \right)$. The third moment or skewness (figure \ref{fig:simQuant}c) becomes non-zero as the steepness increases, and the particle position distribution only remains symmetric (zero skewness) for small steepness. 

Note that when studying drift induced by individual focused wave groups, \citet{Pizzo2019} observed a sharp transition from a quadratic dependence of the drift velocity on steepness below the breaking threshold to a linear dependence above the breaking threshold. For individual waves or wave groups there is a clear threshold for the steepness above which breaking occurs. However, because we are considering many irregular waves only some of which break, the number of breaking events and thus the mean drift velocity increase continuously as a function of  $\eps$ as shown in figure \ref{fig:lambda_G}a. 


\section{Conclusions}\label{sec:conclusion}
In this paper we have developed a stochastic framework to describe particle drift in irregular sea states using a jump-diffusion process to model the enhanced drift due to breaking previously observed by \citet{Deike2017}, \citet{Pizzo2019} and \citet{Sinnis2021}. The framework consisting of a stochastic differential equation for particle transport and a corresponding Fokker--Planck equation for the evolution of its probability distribution can be used to predict mean drift and its higher-order statistical moments given basic information describing the sea state, such as its spectrum or summary parameters thereof (i.e., significant wave height and peak period). We compare long-time laboratory experiments with a large number of particles with our theoretical predictions and find good agreement, including specifically for the contribution of the jump process to model enhanced transport by breaking waves. 

For an irregular wave field with negligible amount of breaking, we experimentally verify that the variance of particle position or the single-particle dispersion is proportional to time, confirming  that the assumption of normal diffusion is valid (in contrast to prediction by \cite{Farazmand2019}). Furthermore, we have described the evolution and quantified the uncertainty of particle transport under the influence of wave breaking, by modeling this as a compound Poisson process, where the amplitudes of the jumps follow a Gamma distribution that is parameterized by the wave steepness. We find that taking into account the jumps induced by breaking waves increases the mean drift  and the variance, and introduces a finite skewness into the particle position distribution, whereas for the non-breaking diffusion problem, the distribution remains normal and thus has zero skewness. Particle tracking laboratory experiments corroborate this. Our results for the enhanced mean drift due to breaking waves are in approximate quantitative agreement with  \cite{Pizzo2019}, who show that the enhancement of drift due to breaking can be up to 30\% compared to the Stokes drift in certain cases.

    
In this paper we have considered long-crested (or unidirectional) waves. In the ocean, sea state are almost always directionally spread (or short-crested). We envisage our model can be readily extended to directional seas using the results of \cite{Kenyon1969} for non-breaking waves. Calibrating the jump-diffusion process for directionally spread breaking waves will require new laboratory experiments. 

Looking forward, we believe that the simple stochastic framework we have developed, which is based on stochastic differential equations used widely in financial economics and climate physics, can be an important tool in uncertainty quantification of prediction models in practical settings, including search and rescue or salvage operations and pollution tracking and clean-up efforts.

\section*{Acknowledgements}
We would like to thank Anton de Fockert and Wout Bakker at Deltares and Pieter van der Gaag, Arie van der Vlies and Arno Doorn at Delft University of Technology for their help setting up and conducting the experiment. This research was performed in part through funding by the European Space Agency  (Grant no. 4000136626/21/NL/GLC/my). D.E. acknowledges financial support from the Swiss National Science Foundation (P2GEP2-191480) and the ONR Grants N00014-21-1-2357 and N00014-20-1-2366. T.S.vdB was supported by a Royal Academy of Engineering Research Fellowship.
\section*{Declaration of interests}
The authors report no conflict of interest.

\bibliographystyle{jfm}
\bibliography{partcleBib}


\appendix
\counterwithin{figure}{section}
\section{Particle tracking}\label{app:DataProcessing}
\begin{figure}
    \centering
    \includegraphics[width=65mm]{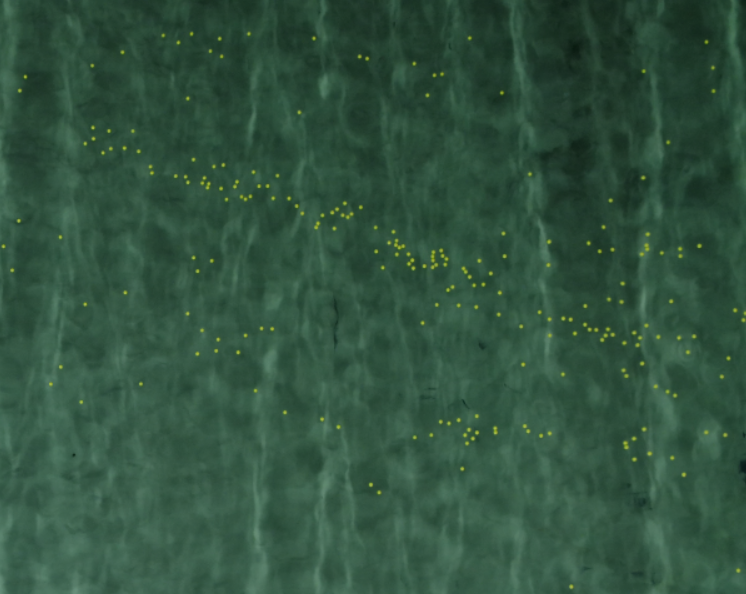}
    \includegraphics[width=65mm]{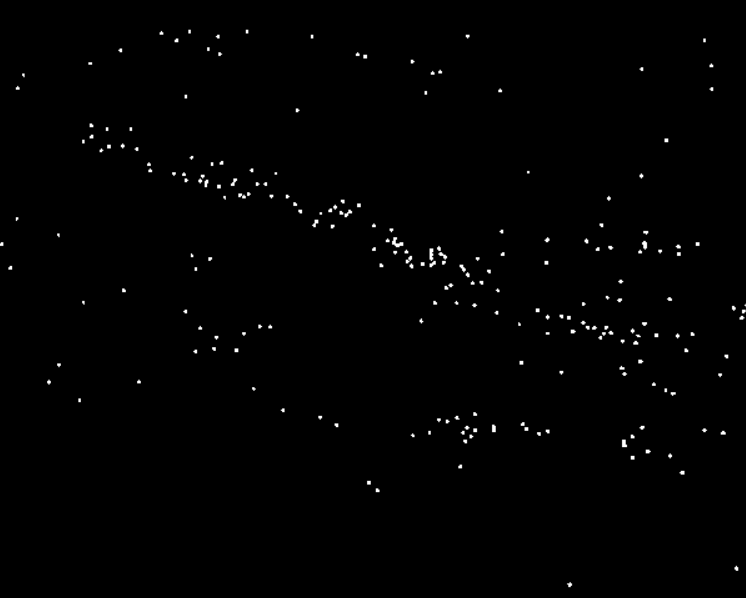}
    \caption{(a) Footage of the Atlantic Basin from the downward-looking camera. The small yellow spheres can be seen floating on the surface. (b) Identification of the spheres after application of the Hue saturation filter in OpenCV.}
    \label{fig:tracking_img}
\end{figure}

A downward-looking camera mounted on a walkway above the basin was used to track particles as they moved across the basin. The camera intriniscis were found by detecting multiple images of a checkerboard in different orientations across the entire field of view (FOV). Figure \ref{fig:tracking_img}(a) shows a top view of the Atlantic Basin with more or less randomly distributed yellow plastic spheres floating on the surface. The spheres are identified and tracked using OpenCV in Python. The image is first filtered using a Hue filter, which produces figure \ref{fig:tracking_img}(b). A Correlation Filter with Channel and Spatial Reliability (CSRT) algorithm was used to track the objects between frames to create trajectories in sub-pixel locations and time. The trajectories were then transformed to the still-water plane, defined by an image of a floating checkerboard, and tank  coordinates ($x,y$) by detecting and inverting the camera intrinsics and applying a measured translation from camera field of view.

The trackers and trajectories were post-processed to eliminate any erroneous tracking events, such as spheres colliding, loss of tracking, or jumps of the particle tracked by the algorithm to a nearby particle, which sometimes occurred when particles were lost momentarily under breaking waves. Finally, the trajectories were manually inspected for quality control.

\section{Jump detection}\label{app:JumpDetection}
\begin{figure}
  \centerline{\includegraphics[width=140mm]{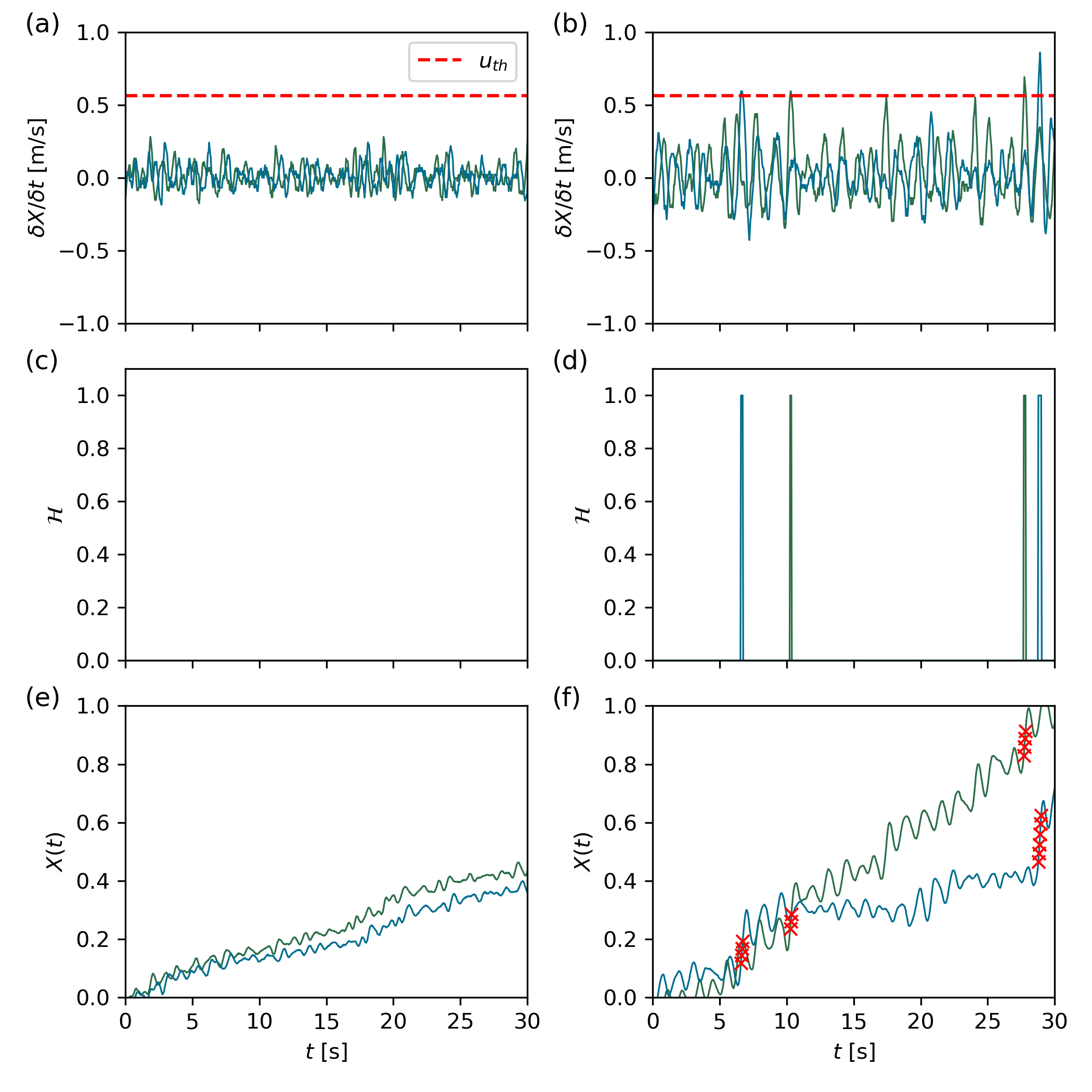}}
  \caption{Jump detection and amplitude estimation process for (\textit{a,c,e}) the least steep waves with $H_s$ = 0.05 m and (\textit{b,d,f}) the steepest waves with $H_s$ = 0.17 m. (\textit{a,b}) Instantaneous particle velocity (before wave averaging) obtained from the derivative of particle position at the camera frame-rate $u_I=\delta X/\delta t$. The dashed line indicates the threshold value $u_{th}=0.3 c$, with $c$ the phase velocity. (\textit{c,d}) The Heaviside function $\mathcal{H}(u_\text{I} (t)-u_{th})$ marks the time segments where the particle is `jumping'. (\textit{e,f}) Particle position $X(t)$ at the camera frame rate, where the red crosses mark the positions for which the Heaviside function is positive and behaviour is classified as `jumping'.}
\label{fig:JumpDet}
\end{figure}

\Cref{fig:JumpDet} displays the steps of the jump detection and jump amplitude (distance travelled in a jump) estimation process for two example trajectories. For $H_s$ = 0.05 m, panel a shows the derivative of particle position over a time-step $\delta t$ determined by the camera frame-rate:  $u_\text{I} = \delta X / \delta t$; we consider this to be the instantaneous velocity (before wave averaging). The dashed line indicates the threshold value $u_{th}=0.3 c$, where $c=\omega/k$ is the phase velocity obtained from the linear dispersion relationship. In panels c and d the Heaviside functions $ \mathcal{H}(u_\text{I} (t)-u_{th})$ mark the time segments where the particle is classified as `jumping'. In panel e and f, based on the Heaviside functions in panels c and d, the jump amplitude $s$ (the distance covered during a jump) can be estimated. 

\section{Drift velocities}\label{app:Velocities}
\begin{figure}
  \centerline{\includegraphics[width=90mm]{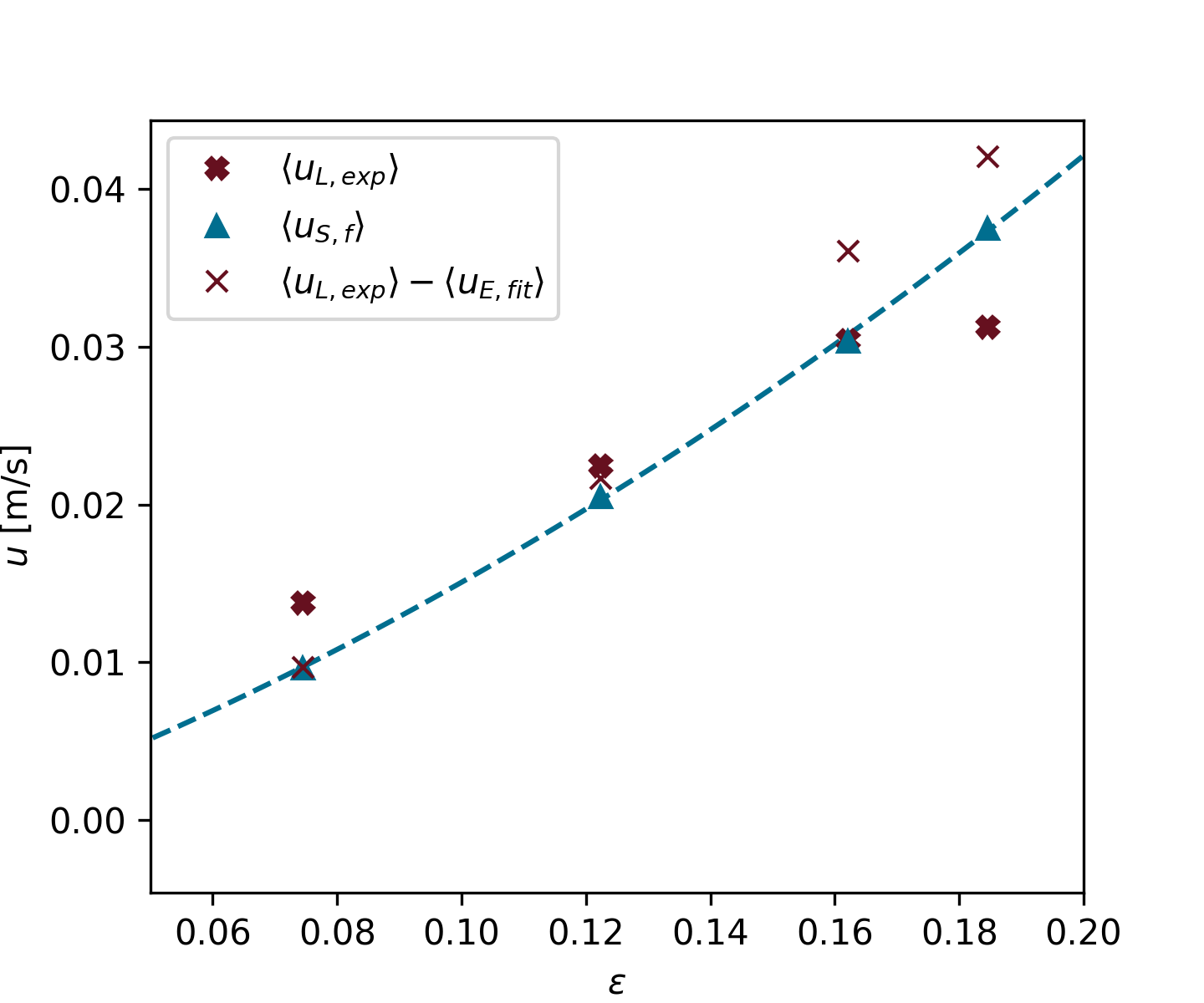}}
  \caption{Drift velocity estimations and measurements as a function of steepness. Stokes drift based on a JONSWAP fit on the spectrum truncated at $\langle u_{\text{S}, f} \rangle$ ($\vartriangle$) and its quadratic fit (dashed line), measured particle drift $\langle u_\text{L,exp} \rangle$ (thick $\times$), measured  particle drift   corrected by model mean flow $\langle u_\text{L,exp} \rangle-\langle u_\text{E,fit} \rangle$ (thin $\times$)}
\label{fig:velocities}
\end{figure}

\Cref{fig:velocities} shows various drift velocity estimations. For our experiments,the Stokes drift is based on a JONSWAP spectrum fitted to the measured spectrum by fitting a JONSWAR spectrum to the experimental spectrum $\langle u_{\text{S}, f} \rangle$, indicated by the blue triangles, with the dashed blue line its quadratic fit. 

We correct the experimentally measured mean drift velocity $\langle u_\text{L,exp} \rangle$ by the mean flow in the model $\langle u_\text{E,fit} \rangle$, resulting in the thin red crosses ($\times$). 
The difference between these and dashed blue line, i.e. the Stokes drift, shows that for low characteristic steepness this coincides with the stokes drift, whereas for high steepness, the stokes drift underestimates the mean drift velocity by a certain fraction. The difference is attributed to the jump process in the model.

\begin{table}
  \begin{center}
\def~{\hphantom{0}}
  \begin{tabular}{lcc}
     $H_s$ &  $\langle u_\text{E, exp} \rangle$  &  $\langle u_\text{E, fit} \rangle$ \\[3pt]
    (m) & (mm/s)   & (mm/s)   \\[3pt]
    0.050 & -2 & +4.1   \\
    0.090 & -10 & +0.8  \\
    0.120 & -6 & -5.6 \\
    0.170 & -17 & -10.8\\
  \end{tabular}
  \caption{Difference between the experimentally measured Eularian mean flow, and its value used in the model.}
  \label{tab:uLdifference}
  \end{center}
\end{table}

\section{Velocity spectrum particles}\label{app:velocityspec}

\begin{figure}
  \centerline{\includegraphics[width=150mm]{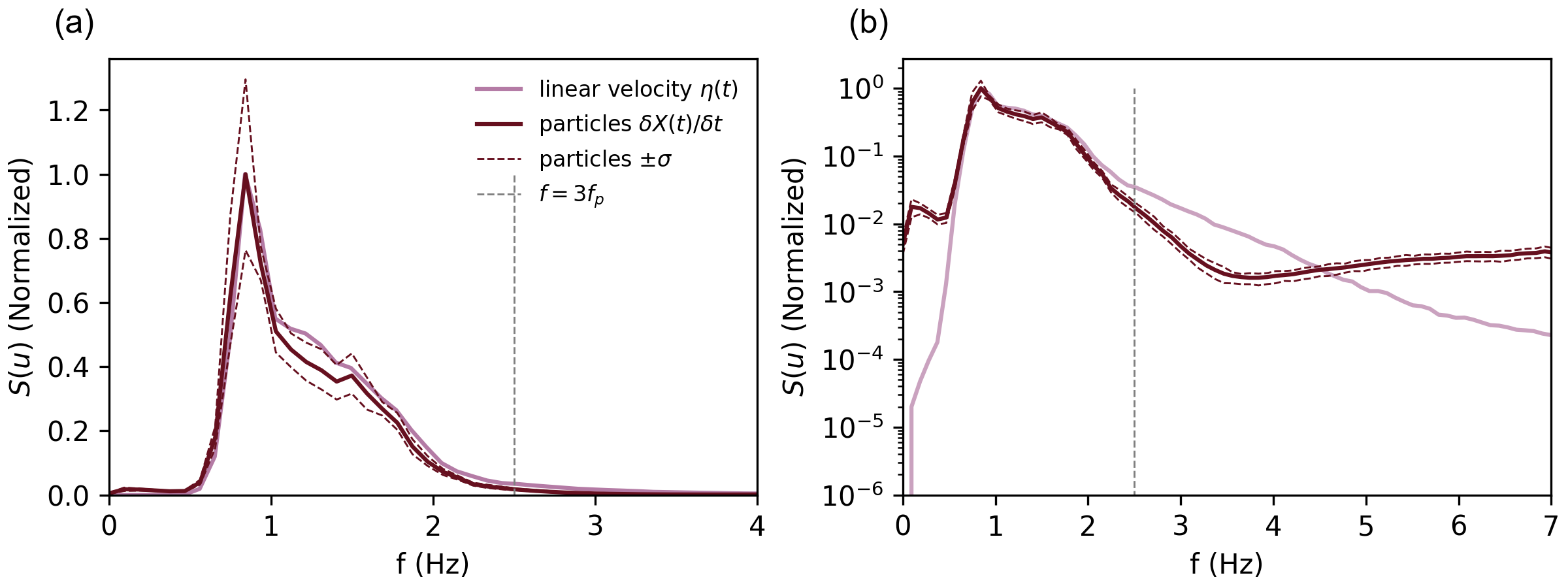}}
  \caption{$H_s$ = 0.05 m. Spectrum of the velocity of the particles (red), dashed lines indicate $\pm 1$ std. The linear velocity calculated from the spectrum of the surface elevation $\eta$ is shown in pink. The gray dashed line indicates $f=3f_0$. (\textit{a}) Linear scale (\textit{b}) Log scale. }
\label{fig:velocityspec}
\end{figure}

\Cref{fig:velocityspec} compares the velocity spectrum of the particle trajectories (dark red) to that of the first order velocity spectrum calculated from the surface elevation (pink). The later has much spectral power for higher frequencies that is not present in the former. Therefore, if this first order velocity does not contribute to the particle movement at these frequencies, the higher order velocity (the Stokes drift) cannot contribute either.

\begin{figure}
  \centerline{\includegraphics[width=150mm]{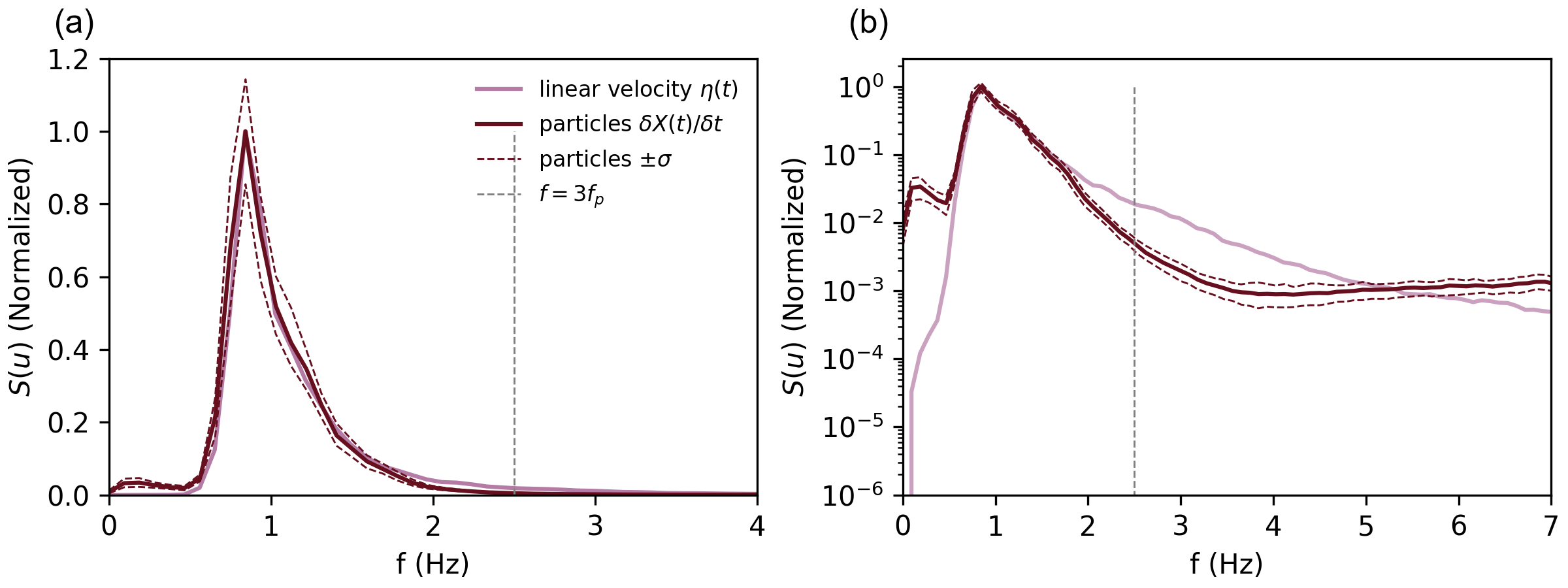}}
  \caption{$H_s$ = 0.09 m. Lines same as in \cref{fig:velocityspec}}
\label{fig:velocityspecHs09}
\end{figure}
\begin{figure}
  \centerline{\includegraphics[width=150mm]{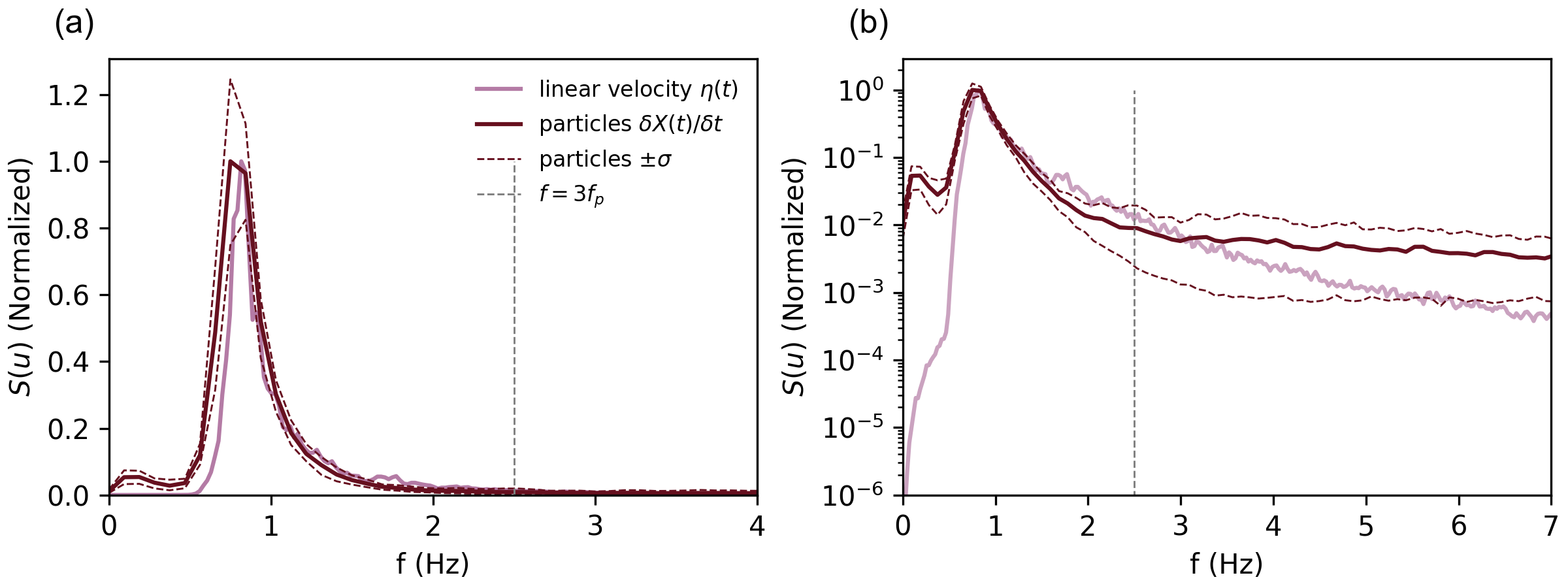}}
  \caption{$H_s$ = 0.12 m. Lines same as in \cref{fig:velocityspec}}
\label{fig:velocityspecHs12}
\end{figure}
\begin{figure}
  \centerline{\includegraphics[width=150mm]{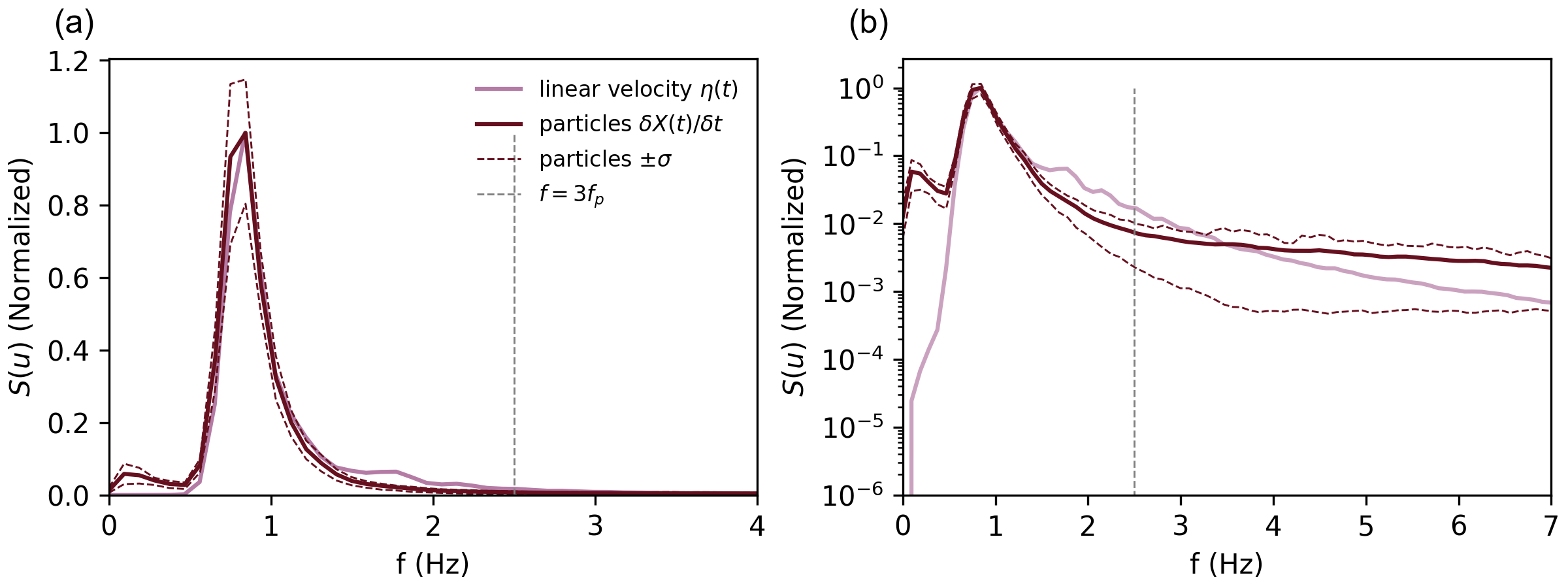}}
  \caption{$H_s$ = 0.17 m. Lines same as in \cref{fig:velocityspec}}
\label{fig:velocityspecHs17}
\end{figure}

\end{document}